\documentclass[12pt]{article}
\usepackage[utf8]{inputenc}
\usepackage{graphicx}
\usepackage{amsmath}
\usepackage{amsfonts}
\usepackage{amssymb}
\usepackage{hyperref}
\usepackage{float}
\usepackage{natbib}
\usepackage{caption}
\usepackage{subcaption}
\usepackage{authblk}
\usepackage{xcolor}
\usepackage{array}

\title{Evaluating and Scoring Ebolavirus Protein-protein Docking Models Using PIsToN}
\author[1]{Azam Shirali}
\author[1]{Vitalii Stebliankin}
\author[1]{Jimeng Shi}
\author[2,3]{Prem Chapagain}
\author[1,3]{Giri Narasimhan\thanks{\tiny Correspondence: Giri Narasimhan, Bioinformatics Research Group (BioRG), Knight Foundation School of Computing and Information Sciences, Florida International University; 11200 SW 8th St, Modesto A. Maidique Campus, Miami, FL 33199, USA. Email: giri@fiu.edu, Ph: (+1) 305-348-3748}}

\affil[1]{\small Bioinformatics Research Group (BioRG), Knight Foundation School of Computing and Information Sciences, Florida International University, 11200 SW 8th St, Miami, 33199, USA.}
\affil[2]{\small Department of Physics, Florida International University, 11200 SW 8th St, Miami, 33199, USA.}
\affil[3]{Biomolecular Sciences Institute, Florida International University, 11200 SW 8th St, Miami, 33199, USA.}

\date{}

\begin{document}

\maketitle

\begin{abstract}

Protein-protein docking is crucial for understanding how proteins interact. Numerous docking tools have been developed to discover possible conformations of two interacting proteins. However, the reliability and success of these docking tools rely on their scoring function. Accurate and efficient scoring functions are necessary to distinguish between native and non-native docking models to ensure the accuracy of a docking tool. Like in other fields where deep learning methods have been successfully utilized, these methods have also introduced innovative scoring functions. An outstanding tool for scoring and differentiating native-like docking models from non-native or incorrect conformations is called Protein binding Interfaces with Transformer Networks (PIsToN). PIsToN significantly outperforms state-of-the-art scoring functions. 
Using models of complexes obtained from binding the Ebola Virus Protein VP40 to the host cell’s Sec24c protein as an example, we show how to evaluate docking models using PIsToN.

\end{abstract}


\section{Introduction}
\label{sec:introduction} 

Protein-protein binding plays a crucial role in many important biological processes and diseases. Most cellular processes, including signaling and immune responses, rely on protein-protein interactions. Undesired protein-protein binding can lead to diseases such as cancer, neurodegenerative disorders, and infectious diseases, including various viral infections. For instance, the Ebola virus uses host cell machinery at every stage of its life cycle—from entry to replication, assembly, and budding. It hijacks cellular pathways for producing viral proteins and assembling new virus particles, while also suppressing the immune response to ensure its survival and spread \cite{vanhook2014ebola}. These processes essentially involve host protein - virus protein interactions. Computational methods such as protein-protein docking (e.g. with HDOCK \cite{yan2020hdock}, ZDOCK \cite{pierce2014zdock})  or multi-protein assembly predictions using AlphaFold 3.0 \cite{abramson2024accurate} can be used to determine the protein-protein complexes to investigate the assembly dynamics and binding mechanisms.  Accurate prediction of the model complexes such as by scoring and ranking docking models is essential for understanding protein function, disease mechanisms, and therapeutic applications. However, developing accurate and efficient scoring functions to differentiate native from non-native structures presents a challenge. Without these scoring functions, the accuracy of docking tools cannot be guaranteed \cite{shirali2025comprehensive}. 

Considerable efforts have been dedicated to developing meaningful scoring functions. This has involved various approaches, such as utilizing force fields and physics features, estimating binding affinity by summing up energy terms before and after bonding, and leveraging knowledge from statistical potentials of known 3D structures. With significant advancements in machine learning (ML) and deep learning (DL), alternative ML-based models have been introduced that can learn to estimate scoring functions and rank docking models. These models are capable of learning complex transfer functions that map a combination of chemical, physical, and geometrical features to predict scoring functions.

Among the state-of-the-art methods, PIsToN (evaluating Protein binding Interfaces with Transformer Networks) uses a trained deep learning model that can differentiate native-like docking models from non-native and incorrect conformations \cite{stebliankin2023evaluating}.
In PIsToN, each protein interface is represented as a set of 2D images. These images represent various geometric, physical, and biochemical properties, capturing atomic-level information of relevant protein characteristics. Additionally, empirical-based energy terms serve as ``hybrid'' inputs to the neural network. 
PIsToN represents binding interfaces as pairs of two-dimensional multi-channel images and then utilizes a trained Vision Transformer (ViT) \cite{dosovitskiy2020image} neural network. It explicitly integrates interaction properties, including atomic distances, relatively accessible surface areas (RASA), van der Waals interactions, complementary surface charges, hydrophobicity values, and more. PIsToN has demonstrated superior performance, significantly outperforming state-of-the-art methods on several well-known datasets \cite{stebliankin2023evaluating}.

The term \textit{``explainability''} refers to the ability of a deep learning model to transparently demonstrate the rationale behind its decisions, making it easier to comprehend why specific outcomes are predicted. In PIsToN, explainability is crucial because it enables users to not only see the score of a docking model but also understand the factors that influenced that score.
The PIsToN architecture uses multi-attention ViT, which helps to emphasize important regions where proteins bind and identify which features (shape, charge, hydropathy, etc.) are most critical to score a model efficiently (see \cite{stebliankin2023evaluating} for more details). Based on our knowledge, PIsToN is the first scoring function to provide explainability, enhancing its trustworthiness for protein docking models.

In this chapter, we will introduce how to use PIsToN as a scoring function for protein-protein docking models and run it on Ebola virus docking models. For this, we take the Ebola virus's structural matrix protein VP40 binding to the host cell's Sec24c protein as an example of host-virus protein-protein binding. The VP40-Sec24c binding aids in transportation to the plasma membrane, ultimately contributing to the development of Ebola virus disease \cite{bhattarai2022ebola}. The VP40 protein of the Ebola virus is able to associate with the plasma membrane of the host cell and has the capacity to form virus-like particles independently \cite{bhattarai2022ebola}. Computational models are used to study the VP40 binding site on Sec24c, which is essential for developing effective vaccines and speeding up the vaccine design process. The organization of this chapter is as follows: Section \ref{sec: material} explains the required inputs and their formats, along with other materials needed for using PIsToN. In Section \ref{sec: methods}, we will guide users through the installation and the process of creating their own docking models. Additionally, in Section \ref{sec: example of usage}, we will demonstrate how to utilize PIsToN on the docking models of the Ebola virus. Important and useful notes are listed in Section \ref{sec: notes} to provide additional information to readers and users.

\section{Materials} \label{sec: material}
\subsection{Input Format} \label{subsec: inputs}

The input to PIsToN is the 3D structure of docking models (in PDB format) of a complex with two interacting proteins. Protein Data Bank (PDB) \cite{berman2000protein} files must have the element symbol of the atoms in columns 77-78. All docking models that need to be scored are required to be listed in a text file, with each complex shown on a separate line in the following format: \emph{PID\_ch1\_ch2}. In this format, \emph{PID} represents the name of the PDB file, while \emph{ch1} and \emph{ch2} represent the names of the chains in the first and second proteins, respectively. 

By running PIsToN, all docking models in the list are scored and saved as an output in \texttt{csv} format. PIsToN assigns a score in the range $[-2, 2]$ to each docking model. Docking models with lower scores are considered to have near-native conformations. Therefore, an ideal binding corresponds to a score of -2, while a score of 2 indicates the most unlikely binding interface and represents a non-native docking model.

\subsection{Data Preprocessing} \label{subsec: preprocessing}
To use PIsToN, we start by preparing and preprocessing the data. For each docking model in PDB format, we create an interface map by projecting surface features from the binding interface of the individual interacting proteins. This results in a multi-channel image, with each channel focused on a specific feature type. Before applying the machine learning model, the data is transformed into 2D images, a method that has been successfully employed in other machine learning applications, such as machine vision.

To generate feature maps, we followed the following steps: 

\begin{enumerate}
    \item First, we refined the protein structures to include side-chain flexibility using FireDock (Fast Interaction Refinement in Molecular Docking) \cite{andrusier2007firedock}. We then computed the binding free energy terms, such as van der Waals, desolvation, insideness, hydrogen and disulfide bonds, electrostatics, and $\pi$-stacking, cation–$\pi$ and aliphatic interactions of the refined structures.

    \item Next, we cropped the docking models to within a specified distance from the center of interaction as the geometric center of contact points.

    \item The solvent-excluded surface was then triangulated and re-scaled to a granularity of 1 Å using the MaSIF data preparation module \cite{gainza2020deciphering}.

    \item We calculated patches and their associated features for each protein individually on the interface. Each patch consists of vertices on a triangulated protein surface within a defined geodesic distance from the interaction center. Additional features such as shape index, curvature, hydrogen-bond potential, charge, and hydropathy were computed for each surface point on a surface using the MaSIF data preparation module \cite{gainza2020deciphering}. We also computed an image for the patch distance, which is generated from a grid of Euclidean distances between points on the two protein surfaces that project corresponding points on the patch pair. Finally, Relative Accessible Surface Area (RASA) was computed for each patch residue using DSSP v2.3 \cite{touw2015series}.

    \item Finally, we converted patch features into an image with pixel intensities proportional to the feature values. Surface points on patches were projected onto a 2D plane using a multidimensional scaling algorithm \cite{mead1992review}.
    
\end{enumerate}

All these steps were saved in user-specified folders in the output directory (see Section \ref{sec: example of usage} for details).

\section{Methods} \label{sec: methods}
In this section, we will guide readers through all the necessary steps for installing and running PIsToN.

\subsection{Installation} \label{subsec: installation}

\noindent \textbf{Step 1: Obtain the PIsToN code}

First, download the PIsToN code repository. This repository includes all the scripts, configuration files, and resources needed to run the trained model. Use \texttt{git}, a version control system, to download (or ``clone'') the latest version of the repository to your local machine using the command:
\begin{verbatim}
        git clone https://github.com/stebliankin/piston
\end{verbatim}
\noindent The above command creates a new directory named \texttt{piston} in your current working directory and will contain all the files needed to run PIsToN. Next, navigate to the \texttt{piston} directory using \texttt{cd} command as shown below:
\begin{verbatim}
        cd piston
\end{verbatim}
\noindent Finally, ensure that the main script, \texttt{piston}, has the correct permissions to be executed as a program:
\begin{verbatim}
        chmod +x piston
\end{verbatim} \vspace{12pt}
\noindent \textbf{Step 2: Install required Python packages in the \texttt{Conda} environment}


For optimal performance and no package conflicts, a dedicated \texttt{Conda} environment is recommended for installing required packages and their dependencies. 
\texttt{Conda} creates an isolated execution environment for PIsToN. 

Use the following commands to achieve this:
\begin{verbatim}
        conda create -n piston python=3.7
        source activate piston
\end{verbatim}
\noindent The version of Python to be used is \texttt{python=3.7}. The environment is then activated by the \texttt{source} command. 
%
%
\noindent Next, the necessary Python packages are installed using the following command:
\begin{verbatim}
        pip3 install \
            tqdm \
            einops \
            keras-applications==1.0.8 \
            opencv-python==4.5.5.62 \
            pandas \
            torch==1.10.1 \
            biopython --upgrade \
            plotly \
            torchsummary \
            torchsummaryX \
            scipy \
            sklearn \
            matplotlib \
            seaborn \
            ml_collections \
            kaleido \
            -U scikit-learn \
            pdb2sql
\end{verbatim} 
%
To execute PIsToN on specific docking models, MaSIF and FireDock must be installed. 
MaSIF can be installed using the data preparation module given in its GitHub site: \url{https://github.com/LPDI-EPFL/masif}.
FireDock can be installed using the following link: \url{https://www.cs.tau.ac.il//~ppdock/FireDock/download.html}.
\bigskip
%

\noindent \textbf{Alternative Step 2: Install required Python packages using the Singularity container}

For users utilizing an HPC (High-Performance Computing) cluster, using a Singularity container can simplify the setup process and ensure consistent runs of PIsToN across different systems. Download the pre-built Singularity container \texttt{piston.sif}, which includes all necessary dependencies, using the following command:
\begin{verbatim} 
  wget https://users.cs.fiu.edu/~vsteb002/piston_sif/piston.sif
\end{verbatim}
\noindent This container was built with Singularity version 3.5.3 and includes all pre-configured dependencies required to run PIsToN. The detailed Singularity definition file \texttt{piston.def}, which describes the environment setup, can be found in the repository under the \texttt{env} folder. This file is particularly useful to understand or modify the environment configuration.

Successful completion of the above steps prepares PIsToN to assess docking models and score them.
An example of its usage is provided below in the next section.

\section{Using PIsToN to Score protein-protein Docking Models} \label{sec: example of usage}

\subsection{Preparing Docking Models}
In this section, we will use PIsToN to evaluate and rank the docking models obtained from the binding of VP40 to the host cell’s Sec24c protein. We will also describe the results and their significance.

The docking models need to be obtained first, before using PIsToN. This can be achieved, for example, by following the steps below:

\begin{enumerate}
    \item Download the VP40 3D structure from the PDB (PDB ID: \texttt{7jzj})

    \item The PDB structure has missing residues and they need to be filled so that the structure of a continuous sequence is used. A complete sequence without the missing residues can be downloaded from UniProt.Download the complete sequence (in Fasta format) from UniProt ID \texttt{Q05128} (\url{https://www.uniprot.org/uniprotkb/Q05128/entry#sequences}). 
       \item Use the SwissModel \cite{bienert2017swiss} webserver (\url{https://swissmodel.expasy.org/interactive}) to fill in missing residues. For this, copy and paste the sequence obtained from UniProt and search for templates. Once the templates are suggested, use \texttt{7jzj.2.A} as the template to build the VP40 dimer model. Download the modeled PDB file of the VP40 dimer structure.

    \item Repeat the previous steps for human Sec24c. Utilize the sequence from UniProt ID \texttt{P53992}, search for a template, and finally use \url{3eh2.1.A} as the template.

    \item 
    Use the HDOCK \cite{yan2020hdock} server (or any of several freely available docking servers) to generate docking models for the complex involving VP40 and Sec24c (in PDB format). 
    HDOCK can generate 100 docking models and can rank them based on their quality. We suggest downloading the top 10 docking models.
    
    \item Finally, use PIsToN to score the top docking models as described below.  
\end{enumerate}
\counterwithout{figure}{section}

\subsection{Evaluating Docking Models} 

Having prepared the docking models, we execute PIsToN for evaluating the binding interfaces and ranking models as described below.

\begin{enumerate}

    \item Change the current directory to where PIsToN is downloaded and create a new folder named \texttt{Models}. Move the docking models to this folder. 

    \item Navigate to this directory by \texttt{cd /path/to/Models}

    \item  List all the files containing the docking models in a text file named \texttt{models\_list.txt}. Each line in this file contains the name of a file with one docking model and specifies the chains of the two proteins separated by an underscore (\texttt{\_}) (see Section \ref{sec: material}). For example, the text lines may look like this:
    \begin{verbatim} 
    model-1_AB_C
    model-2_AB_C
    model-3_AB_C
    \end{verbatim}
    Each of the three docking models in the example above contains the VP40 dimer with chains A and B and the Sec24c protein with chain A. We renamed the Sec24c protein chain to chain C to ensure distinct chain names used in the two proteins. 
    
    \item Run the Python script below with the specified arguments:
    \begin{verbatim} 
    python ../piston.py infer --pdb_dir ./ 
                               --list ./models_list.txt 
                               --out_dir ./piston_out
    
    \end{verbatim}
    The \texttt{infer} module calculates PIsToN scores for docking models and generates visual representations of the associated interface maps.

    Option \texttt{--pdb\_dir} specifies the path to the PDB files of the docking models.

    Option \texttt{--list} specifies the file with the list of docking models.
  
    Option \texttt{--out\_dir} is the directory where the output files will be saved.
  
    \item The output generated by PIsToN is organized into several files.
    
    \texttt{PIsToN\_scores.csv}: Contains PIsToN scores in \texttt{csv} format, where lower scores indicate better binding. Table \ref{table:scores} shows the scores (rounded to 4 decimal places) obtained for the Ebola virus docking models. See section \ref{sec: material} for details. 
    \begin{table}[ht]
    \caption{PIsToN scores for Ebola virus docking models} \label{table:scores}
    \centering
    \begin{tabular}{>{\raggedright\arraybackslash}p{3.4cm} >{\centering\arraybackslash}p{2cm}}
    \hline
    \textbf{Docking Model} & \textbf{Score} \\ \hline
    model-2 & -0.744 \\ 
    model-4 & -0.540 \\ 
    model-7 & 0.009 \\ 
    model-10 & 0.022 \\ 
    model-1 & 0.452 \\ 
    model-9 & 0.508 \\ 
    model-3 & 0.633 \\ 
    model-8 & 0.864 \\ 
    model-5 & 0.941 \\ 
    model-6 & 0.999 \\ 
    \hline
    \end{tabular}
    \end{table} 
    
     These scores represent PIsToN's predictions of the strength of the interaction between the two proteins in the complex, a crucial piece in our understanding of the complex. Docking models that obtained a score less than zero are classified as near-native and correct models, while those that obtained scores more than zero are classified as non-native and incorrect models.
 
    \texttt{gird\_16R}: A directory containing interface maps in \texttt{NumPy} format.

    \texttt{intermediate\_files}: This folder contains the intermediate files created by PIsToN, including protein structures after various processing steps such as protonation, refinement, cropping, and the many patches of interest.

    \texttt{patch\_vis}: This folder contains \texttt{html} files with interactive visualizations of interface maps. 
\end{enumerate}

\vspace{3mm}

\begin{figure}[htbp!] 
    \centering
    \includegraphics[width=1\textwidth]{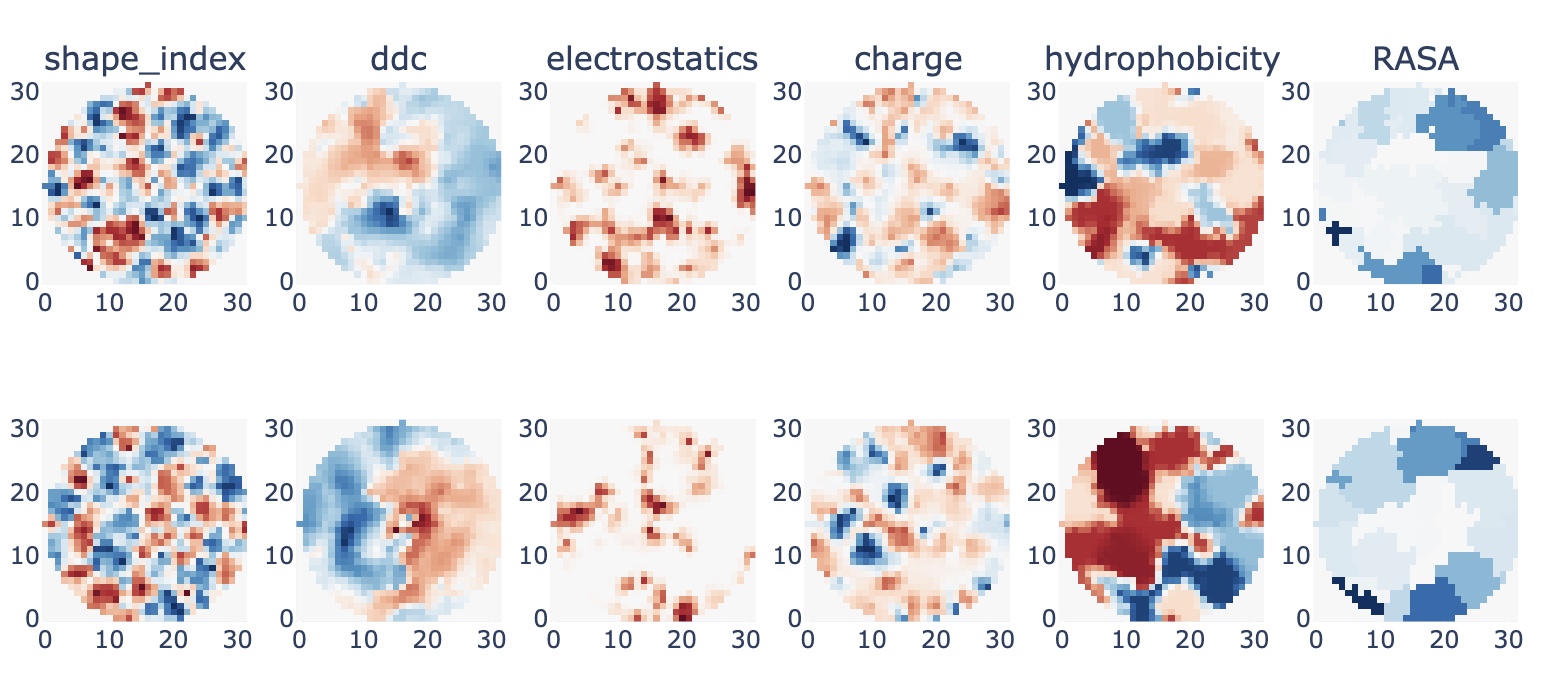}
    \caption{ The interactive patch pair for \texttt{model-2} using its \texttt{html} file in \texttt{patch\_vis} folder.} \label{fig:fea-maps}   
\end{figure}

As \texttt{model-2} is ranked as the best model by PIsToN among all 10 docking models, we visualized its interface maps in Figure \ref{fig:fea-maps}.)

\begin{figure}[htbp!] 
\centering
\includegraphics[width=0.9\textwidth]{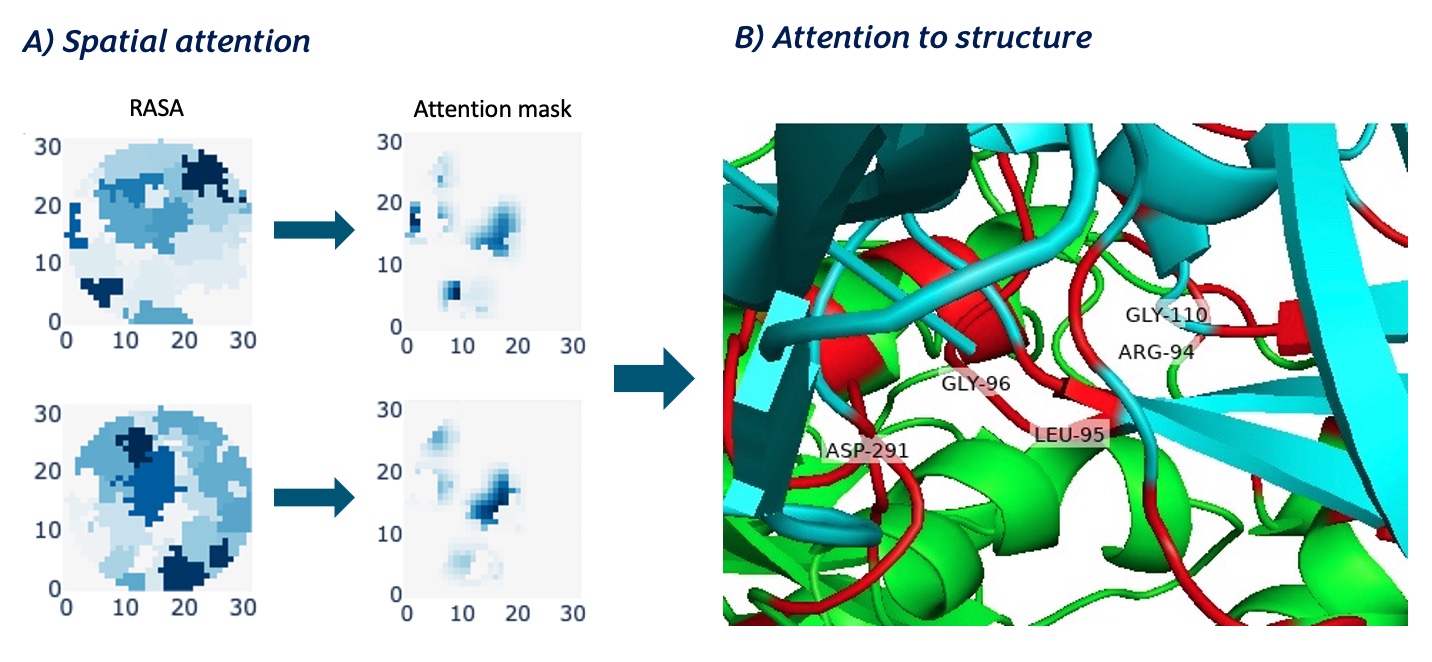}
\caption{ \textbf{A)} Identifying significant features and corresponding significant pixels from the spatial attention map for \texttt{model-2}. The color of each pixel signifies the value
of the corresponding feature on a patch. \textbf{B)} Structure of \texttt{model-2} at the interface. Significant residues are highlighted as a red strand. 
The residues of interest in the binding are ARG-94, LEU-95, GLY-96, GLY-110, and ASP-291.}  \label{fig:attention}  
\end{figure}

PIsToN helps us to analyze the essential features and binding sites that led to a classification decision. Figure \ref{fig:attention} is drawn to identify significant features contributing the most, where PIsToN identified it as a correct model. For model-2 PIsToN paid the greatest attention to RASA when predicting it as a near-native model. PIsToN can help to identify the significant pixels and find the region of significance, which for model-2 was close to the interface. A portion of the structure of \texttt{model-2} at the interface is depicted in Figure \ref{fig:attention}-B, with significant residues highlighted by PIsToN.

\section{Notes}\label{sec: notes}

\begin{itemize}
    \item The complex of Ebola virus matrix protein VP40 and human Sec24c was used as an example to show how to run PIsToN to assess protein-protein docking models.


    \item Note that the PDB files of docking models must contain the element symbol of the atoms in columns 77-78.  

    \item The docking models cannot contain any underscores (\_) in their filenames.

    \item The same procedure can be followed for other protein-protein complexes.

    \item More details on using PIsToN can be found here:  \url{https://github.com/stebliankin/piston/tree/main}
\end{itemize}


\bibliographystyle{unsrt}
\bibliography{reference}

\end{document}